\documentclass[english]{IEEEtran}
\usepackage{amsmath,balance}
\usepackage[T1]{fontenc}
\usepackage[latin9]{inputenc}
\usepackage{amsthm}
\usepackage{amsmath}
\usepackage{amssymb}
\usepackage{graphicx}
\usepackage{balance}
\usepackage{epstopdf}
\usepackage{algpseudocode}
\usepackage{algorithm}
\usepackage{bbold}

\makeatletter
\theoremstyle{plain}
\newtheorem{thm}{\protect\theoremname}
\theoremstyle{remark}
\newtheorem{rem}[thm]{\protect\remarkname}

\makeatother

\usepackage{babel}
\providecommand{\remarkname}{Remark}
\providecommand{\theoremname}{Theorem}
\graphicspath{{Figures/}}

\begin{document}

\title{On Distributed Frequency Estimation in Three-Phase Power Distribution Networks}

\author{Sayed Pouria Talebi \\Department of Electrical and Electronic Engineering, Imperial College London, SW7 2AZ, U.K. E-mail:~s.talebi12@imperial.ac.uk.}
\maketitle
\begin{abstract}
The earlier work of the author on Frequency estimation in three-phase power systems in \cite{ME} is expanded to the distributed setting in order present a framework for the implementation of such a frequency estimator in real-world power distribution networks. For rigor,  the mean and mean square performance of the distributed frequency estimator is analyzed. The performance of the developed algorithm is validated through simulations on both synthetic data and real-world data recordings, where it is shown to outperform standard linear and the recently introduced widely liner frequency estimators. 
\end{abstract}
 

\section{Introduction}

The power grid is designed to operate optimally at a nominal frequency and in a balanced fashion \cite{key-1}. Large deviations from the nominal frequency, which typically occur as a result of mismatch between power generation and consumption, have adverse effects on the performance of different components of the grid, such as compensators and loads \cite{key-2}; thus, making frequency stability as one of the most important factors in power quality \cite{key-3}. Therefore, accurate frequency estimation is a prerequisite to establishing frequency stability in the grid and ensuring power quality.

The need for accurate frequency estimation in power grids is even more profound when considering current trends in smart grid technology that incorporate distributed power generation based on renewable energy sources. In this setting, the wide-area grid is divided into a number of self contained sections called micro-grids, with some micro-grids becoming independent in power generation and disconnecting from the wide-area grid for prolonged lengths of time referred to as islanding. Perfect synchrony is required to connect micro-grids and manage islanding; consequently, many smart grid control and management techniques are dependent on accurate estimation of frequency under both balanced and unbalanced operating conditions \cite{key-4}.

The importance of frequency estimation in power grids has motivated the introduction of a variety of algorithms for this purpose, including phase-locked loops (PLL) \cite{key-5}-\cite{key-6}, recursive Newton-type frequency estimation algorithms \cite{key-7}, Fourier transform-based methods \cite{key-8}, state space frequency estimation algorithms established on the Kalman and extended Kalman filters \cite{key-9}-\cite{key-10}, and adaptive notch filter for direct estimation of frequency and its rate of change \cite{key-11}. However, these techniques are either based on using the information of a single phase and cannot fully characterize three-phase systems \cite{key-12}, especially during crucial moments where one or two of the phases encounter a sudden drop in voltage or short circuit referred to as voltage sags \cite{key-13}-\cite{key-15}, or are based on standard complex linear models that are shown to experience large oscillatory errors at  twice the frequency of the system when the three-phase system is unbalanced \cite{key-11},\cite{key-16}. 

In order to introduce a robust frequency estimator for both balanced an unbalanced power systems, the Clarke transform and widely linear modeling of complex-valued signals have been used in \cite{key-17}, where an algorithm based on the augmented least mean square (ACLMS) adaptive filter has been presented. The Clarke transform and widely linear modeling have also been used in  \cite{key-18} to present a frequency estimator based on the augmented complex Kalman filter (ACKF) that outperforms ACLMS based methods.  

An important development in smart grid technology is the recent introduction communication standards that allow measurement units to exchange information with their neighboring  units over the power grid without the need for a dedicated communication infrastructure \cite{key-19}. Although frequency should be estimated locally, the ability to share information with neighboring nodes can be explored to enhance the performance of frequency estimators specially in small networks, such as micro-grids. A number of distributed signal processing strategies based on the LMS \cite{key-20}-\cite{key-21}, ACLMS \cite{key-22}, and Kalman \cite{key-23} filtering algorithms have been introduced; furthermore, a frequency estimator for three-phase power distribution networks based on the diffusion-ACLMS that employs single-hop communication has been presented in \cite{key-24}. However, these distributed estimation algorithms do not account for the low average number of connections per node in power distribution networks and consider all nodes of the network to be suffering from the same voltage sag.

A robust frequency estimator for three-phase power systems has been developed based on the complex valued widely linear adaptive filtering by the author and his colleges in~\cite{ME}. In this work, the frame work has been expanded to the distributed setting to address the implementation of such a frequency estimator in power distribution networks tacking into account practical considerations raised in this section. For rigor, the contribution of the distributed estimation strategy to the mean and mean-squared error performance of the developed algorithm is analyzed. Finally, the concepts are verified using simulations on both synthetic and real-world data recordings.  

\section{Background}

\subsection{Widely linear estimation}

For a random variable, $\mathbf{x}$, the standard covariance, $E[\mathbf{x}\mathbf{x}^{H}]$, is widely considered as the second-order information measure; however, the standard covariance is only adequate for second-order circular (proper) complex random valuables \cite{key-25}. The full description of the second-order information of a general complex random variable is only possible through the augmented complex statistics, where the complex random variable, $\mathbf{x}$, is augmented with its conjugate, $\mathbf{x}^{*}$, to give the augmented random variable as $\mathbf{x}^{a}=[\mathbf{x}^{T},\mathbf{x}^{H}]^{T}$. The augmented covariance matrix can now be expressed as
\[
\mathbf{C}^{a}_{\mathbf{x}}=E[\mathbf{x}^{a}\mathbf{x}^{aH}]=
\begin{bmatrix}
\mathbf{C}_{\mathbf{x}} & \mathbf{R}_{\mathbf{x}}
\\
\mathbf{R}^{H}_{\mathbf{x}} & \mathbf{C}^{H}_{\mathbf{x}}
\end{bmatrix}
\]
where $E[\cdot]$ represents the statistical expectation, while $\mathbf{C}_{\mathbf{x}}=E[\mathbf{x}\mathbf{x}^{H}]$ is the standard covariance and $\mathbf{R}_{\mathbf{x}}=E[\mathbf{x}\mathbf{x}^{T}]$ is the pseudo-covariance \cite{key-25}. Second-order circular random variables, for which the probability distribution is rotation invariant, have a vanishing pseudo-covariance; however, for general complex random variables both the covariance and pseudo-covariance are required to fully exploit their second-order statistics \cite{key-25}. 

To introduce an optimal second-order estimator for the generality of complex-valued signals, first consider the real-valued minimum mean square error (MMSE) estimator that estimates $y$ conditional to observation $x$, given by 
\[
\hat{y}=E[y|x]
\]
where $\hat{y}$ is the estimate of $y$. For zero-mean and jointly Gaussian $x$ and $y$ the optimal solution is the strictly linear estimator given by
\[
\hat{y}=\mathbf{h}^{T}\mathbf{x}
\] 
where $\mathbf{h}$ is a vector of coefficients and $\mathbf{x}$ is a vector of passed observations (regressor). For complex-valued random variables the MMSE estimator should be expressed in terms of the real and imaginary components \cite{key-26}, which yields
\begin{equation}
\hat{y}=E[y_{r}|x_{r},x_{j}]+jE[y_{j}|x_{r},x_{j}].
\label{eq:MMSE-real-imaginary}
\end{equation}
Replacing $x_{r}=(x+x^{*})/2$ and $x_{j}=j(x^{*}-x)/2$ into the above expression gives
\[
\hat{y}=E[y_{r}|x,x^{*}]+jE[y_{j}|x,x^{*}].
\]
Therefore, the optimal MMSE estimator for complex-valued zero-mean and jointly Gaussian $x$ and $y$ becomes
\begin{equation}
\hat{y} = \mathbf{h}^{T}\mathbf{x}+\mathbf{g}^{T}\mathbf{x}^{*}
\label{eq:widely linear estimator}
\end{equation}
which can be more elegantly presented as
\[
\underbrace{\begin{bmatrix}\hat{y} \\ \hat{y}^{*}\end{bmatrix}}_{\hat{\mathbf{y}}^{a}}=
\underbrace{\begin{bmatrix}\mathbf{h}  & \mathbf{g} \\ \mathbf{g}^{*} & \mathbf{h}^{*} \end{bmatrix}}_{\mathbf{W}^{a}}
\underbrace{\begin{bmatrix}\mathbf{x} \\ \mathbf{x}^{*} \end{bmatrix}}_{\mathbf{x}^{a}}
\] 
where $\hat{\mathbf{y}}^{a}$ and $\mathbf{x}^{a}$ are the augmented estimation and augmented regressor vectors, while $\mathbf{W}^{a}$ is the augmented weight matrix. The estimator in (\ref{eq:widely linear estimator}) is linear in both $\mathbf{x}$ and $\mathbf{x}^{*}$; therefore, it is referred to as the widely linear estimator.

The concept of augmented complex statistics and widely linear estimation have been exploited in \cite{key-27} to introduce a  class of ACKF, including the augmented complex extended Kalman filter (ACEKF). The state evolution and observation equations of the ACEKF are given by
\[
\begin{aligned}
\mathbf{x}^{a}_{k}=&f^{a}(\mathbf{x}^{a}_{k-1})+ \mathbf{u}^{a}_{k}
\\
\mathbf{y}^{a}_{k}=&g^{a}(\mathbf{x}^{a}_{k}) + \mathbf{n}^{a}_{k}
\end{aligned}
\]   
where $f^{a}(\cdot)$ is the augmented state evolution function, $g^{a}(\cdot)$ is the augmented observation function, whereas $\mathbf{u}^{a}_{k}$ and $\mathbf{n}^{a}_{k}$ are the augmented state evolution and observation noise vectors, while $\mathbf{x}^{a}_{k}$ and $\mathbf{y}^{a}_{k}$ are the augmented state and observation vectors. The operations of the ACEKF are summarized in Algorithm-\ref{Al:ACEKF}, where $\mathbf{A}^{a}_{k}$ and $\mathbf{H}^{a}_{k}$ represent the Jacobian matrix of the state evolution and observation functions at time instant $k$, whereas $\mathbf{C}^{a}_{\mathbf{u},k}$ and $\mathbf{C}^{a}_{\mathbf{n},k}$ represent the augmented state evolution and observation noise covariance matrices \cite{key-27}.    

\begin{algorithm}
\caption{ACEKF}
Initialize: $\hat{\mathbf{x}}^{a}_{0|0}$ and $\hat{\mathbf{M}}^{a}_{0|0}$ \\ \vspace{5pt}
For $k=1,2,...$: \\ \vspace{5pt}
$\hat{\mathbf{x}}^{a}_{k|k-1}=f^{a}(\hat{\mathbf{x}}^{a}_{k-1|k-1})$ \\  \vspace{5pt}
$\hat{\mathbf{M}}^{a}_{k|k-1}=\mathbf{A}^{a}_{k}\hat{\mathbf{M}}^{a}_{k-1|k-1}\mathbf{A}^{aH}_{k}+\mathbf{C}^{a}_{\mathbf{u},k}$ \\  \vspace{5pt}
$\mathbf{G}^{a}_{k}=\hat{\mathbf{M}}^{a}_{k|k-1}\mathbf{H}^{aH}_{k}\big(\mathbf{H}_{k}\hat{\mathbf{M}}^{a}_{k|k-1}\mathbf{H}^{aH}_{k}+\mathbf{C}^{a}_{\mathbf{n},k}\big)^{-1}$ \\  \vspace{5pt}
$\hat{\mathbf{x}}^{a}_{k|k} = \hat{\mathbf{x}}^{a}_{k|k-1} + \mathbf{G}^{a}_{k}\big(\mathbf{y}^{a}_{k}-g^{a}(\hat{\mathbf{x}}^{a}_{k|k-1})\big)$ \\  \vspace{5pt}
$\hat{\mathbf{M}}^{a}_{k|k}=\left(\mathbf{I}-\mathbf{G}^{a}_{k}\mathbf{H}^{a}_{k}\right)\hat{\mathbf{M}}^{a}_{k|k-1}$
\label{Al:ACEKF}
\end{algorithm}

\subsection{Three-phase power systems}

The instantaneous voltages of each phase in a three-phase power system are given by \cite{key-28}
\[
\begin{aligned}
v_{a,k}=&V_{a,k}\text{cos}(2\pi f\Delta Tk+\phi_{a,k})
\\
v_{b,k}=&V_{b,k}\text{cos}\big(2\pi f\Delta Tk+\phi_{b,k}+\frac{2\pi}{3}\big)
\\
v_{c,k}=&V_{c,k}\text{cos}\big(2\pi f\Delta Tk+\phi_{c,k}+\frac{4\pi}{3}\big)
\end{aligned}
\]
where $V_{a,k}$, $V_{b,k}$, and $V_{c,k}$ are instantaneous amplitudes, $\phi_{a,k}$, $\phi_{b,k}$, and $\phi_{c,k}$ are instantaneous phases, $f$ is the system frequency, and $\Delta T=1/f_{s}$  is the sampling interval with $f_s$ denoting the sampling frequency. The Clarke transform, given by \cite{key-28}
\[
\left[\begin{array}{c} v_{0,k}\\ v_{\alpha,k}\\ v_{\beta,k}
\end{array}\right]=\sqrt{\frac{2}{3}}\left[\begin{array}{ccc}
\frac{\sqrt{2}}{2} & \frac{\sqrt{2}}{2} & \frac{\sqrt{2}}{2}\\
1 & -\frac{1}{2} & -\frac{1}{2}\\
0 & \frac{\sqrt{3}}{2} & -\frac{\sqrt{3}}{2}
\end{array}\right]\left[\begin{array}{c}
v_{a,k}\\
v_{b,k}\\
v_{c,k}
\end{array}\right]
 \]
maps the three-phase power system onto a new domain where they are represented by $v_{k}=v_{\alpha,k}+jv_{\beta,k}$ while in most practical application $v_{0,k}$ is ignored and only serves the role of making the Clarke transform reversible. 

In a balanced three-phase system, $V_{k}=V_{a,k}=V_{b,k}=V_{c,k}$ and $\phi_{k}=\phi_{a,k}=\phi_{b,k}=\phi_{c,k}$; therefore, $v_{0,k}=0$ resulting in
\begin{equation}
v_k=\sqrt{\frac{3}{2}}V_{k}e^{j(2 \pi f \Delta T k + \phi_{k})}
\label{eq:balanced}
\end{equation}
which can be expressed by employing the first order linear autoregressive model
\[
v_{k}=e^{j2 \pi \Delta T}v_{k-1}
\]
where the term $e^{j2 \pi \Delta T}$ is referred to as the phase increment. 

The expression in (\ref{eq:balanced}) shows that when the three-phase power system is balanced, $v_{k}$ is consisted of only a positive sequenced element; hence, it will trace a circle on the complex plane making the distribution of $v_{k}$  rotation invariant (complex circular) \cite{key-17}-\cite{key-18}. Moreover, under balanced operating condition the frequency of the system can be estimated by standard linear complex Kalman filters employing the state space model given in Algorithm-\ref{Al:L-SS}, where $x_k=e^{j2 \pi \Delta T}$ is the phase increment \cite{key-10}.
\begin{algorithm}
\caption{Linear state space model (L-SS)}
State evolution equation: $\begin{bmatrix}x_{k}\\ v_{k}\end{bmatrix}=\begin{bmatrix}x_{k-1}\\ x_{k-1}v_{k-1}\end{bmatrix}+\mathbf{u}_{k}$\\ \vspace{10pt}
Observation equation: $v_k=\begin{bmatrix}0&1\end{bmatrix}\begin{bmatrix}x_{k}\\ v_{k}\end{bmatrix}+n_{k}$ \\  \vspace{5pt}
Estimate of frequency: $\hat{f}_{k}=\frac{1}{2 \pi \Delta T} \Im \left( \text{ln}\left(x_k \right) \right)$
\label{Al:L-SS}
\end{algorithm}

In practice, a wide range of phenomena, such as voltage sags, load imbalance, and faults in the transmission line,  will lead to unbalanced operating conditions in three-phase power systems \cite{key-13}-\cite{key-14}. Under unbalanced operating conditions \cite{key-17}
\[
v_k=A_{k}e^{j(2 \pi f \Delta T k +\phi_{k})} + B_{k}e^{-j(2 \pi f \Delta T k + \phi_{k})}
\]
where
\[
\begin{aligned}
A_{k}=&\frac{\sqrt{6}\left(  V_{a,k} + V_{b,k} + V_{c,k} \right)}{6}
\\
B_{k}=&\frac{\sqrt{6}\left(  2V_{a,k} - V_{b,k} - V_{c,k} \right)}{12} - j\frac{\sqrt{2}\left(   V_{b,k} - V_{c,k} \right)}{4} \cdot
\end{aligned}
\]
and all phase shifts were considered to be equal to $\phi_{k}$. Therefore, $v_{k}$ comprises both a positive and a negative sequenced element and will trace an ellipse in the complex plane, making the distribution of $v_k$ non-circular.

In order to accommodate both balanced and unbalanced systems, it has been shown that $v_{k}$ can be expressed by employing the first order widely linear autoregressive model
\[
v_{k}=h_{k-1}v_{k-1}+g_{k-1}v^{*}_{k-1}
\]
where $h_{k}$ and $g_{k}$ are the linear and conjugate weights respectively \cite{key-17}. The fundamental frequency of both balanced and unbalanced three-phase power systems can now be estimated by a ACEKF employing the state space model given in Algorithm-\ref{Al:WL-SS} \cite{key-18}.
\begin{algorithm}
\caption{Widely linear state space model (WL-SS)}
State evolution equation: \[\begin{bmatrix}h_{k}\\g_{k}\\ v_{k} \\h^{*}_{k}\\g^{*}_{k}\\ v^{*}_{k}\end{bmatrix}=\begin{bmatrix}h_{k-1}\\g_{k-1}\\h_{k-1} v_{k-1}+g_{k-1} v^{*}_{k-1}\\h^{*}_{k-1}\\g^{*}_{k-1}\\h^{*}_{k-1} v^{*}_{k-1}+g^{*}_{k-1} v_{k-1}\end{bmatrix}+\mathbf{u}^a_{k}\] \\ \vspace{5pt}
Observation equation: \[ \begin{bmatrix}v_{k}\\v^{*}_{k}\end{bmatrix}=\begin{bmatrix}
0 & 0 & 1 & 0 & 0 & 0\\ 
0 & 0 & 0 & 0 & 0 & 1
\end{bmatrix}\begin{bmatrix}h_{k}\\g_{k}\\ v_{k} \\h^{*}_{k}\\g^{*}_{k}\\ v^{*}_{k}\end{bmatrix}+\mathbf{n}^{a}_k \] \\  \vspace{5pt}
Estimate of frequency: 
$\hat{f}_{k}=\frac{1}{2 \pi \Delta T}\text{arcsin}\left(\Im\left(h_k + a_k\right)\right)$\\  \vspace{5pt}
where \[a_k=-j\Im\left(h_k\right)+j\sqrt{\Im^2\left(h_k\right)-\left|g_k\right|^2}\]
\label{Al:WL-SS}
\end{algorithm}

\begin{rem}
Observe that in Algorithm-\ref{Al:WL-SS} the system frequency is calculated as a function of the states. This significantly increases the computational complexity of the algorithm  and can have a detrimental effect on its performance.
\end{rem}  

For a general three-phase system $v_{k}$ can be expressed as \cite{ME} 
\[
v_{k} = \Lambda_{I,k}\text{cos}(2\pi f \Delta T k) - \Lambda_{Q,k}\text{sin}(2 \pi f \Delta T k)
\]
where
\[
\begin{aligned}
\Lambda_{I,k}= & \sqrt{\frac{2}{3}}V_{a,k}\text{cos}(\phi_{a,k})+\big( \frac{j\sqrt{3}-1}{\sqrt{6}}\big)V_{b,k}\text{cos}\big(\phi_{b,k}+\frac{2\pi}{3}\big)
\\
&-\big( \frac{j\sqrt{3}+1}{\sqrt{6}}\big)V_{c,k}\text{cos}\big(\phi_{c,k}+\frac{4\pi}{3}\big)
\\
\Lambda_{Q,k}= & \sqrt{\frac{2}{3}}V_{a,k}\text{sin}(\phi_{a,k})+\big( \frac{j\sqrt{3}-1}{\sqrt{6}}\big)V_{b,k}\text{sin}\big(\phi_{b,k}+\frac{2\pi}{3}\big)
\\
&-\big( \frac{j\sqrt{3}+1}{\sqrt{6}}\big)V_{c,k}\text{sin}\big(\phi_{c,k}+\frac{4\pi}{3}\big).
\end{aligned}
\]
Replacing the $\text{sin}(\cdot)$ and $\text{cos}(\cdot)$ with their polar representations yields
\[
v_{k}=\underbrace{\big(\frac{\Lambda_{I,K}}{2}-\frac{\Lambda_{Q,k}}{2}\big)e^{j 2 \pi f \Delta T}}_{v^{+}_{k}} + \underbrace{\big(\frac{\Lambda_{I,K}}{2}+\frac{\Lambda_{Q,k}}{2}\big)e^{-j 2 \pi f \Delta T}}_{v^{-}_{k}}
\]
where $v_{k}$ has been separated into two counter rotating elements, $v^{+}_{k}$ with only a positive  and $v^{-}_{k}$ with only a negative sequenced element \cite{ME}. The two counter rotating elements can be modeled individually by employing the linear autoregressive models 
\begin{equation}
v^{+}_{k}=e^{j 2 \pi f \Delta T}v^{+}_{k-1} \text{ and }v^{-}_{k}=e^{-j 2 \pi f \Delta T}v^{-}_{k-1}
\label{eq:linear regressive model}
\end{equation}
where the phase increments of the positive and negative sequenced elements are complex conjugates of each other. Therefore, $v_k$ can be expressed using the widely linear autoregressive model given by
\begin{equation}
\begin{bmatrix} v_{k} \\ v^{*}_{k} \end{bmatrix} = \begin{bmatrix} v^{+}_{k-1} & v^{-}_{k-1} \\ v^{-*}_{k-1} & v^{+*}_{k-1} \end{bmatrix} \begin{bmatrix} x_{k} \\ x^{*}_{k} \end{bmatrix}
\label{eq:widely linear model of v}
\end{equation}
where $x_{k}=e^{j 2 \pi f \Delta T}$ and represents the phase increment \cite{ME}.

Taking into account  the widely linear autoregressive model in (\ref{eq:widely linear model of v}), the frequency of the three-phase power system can be estimated by a ACEKF employing the widely linear state space model presented in Algorithm-\ref{Al:N-SS}, where the fundamental frequency of the system is directly estimated from the phase increment, which is modeled as a state \cite{ME}. 

\begin{algorithm}
\caption{The new state space model (N-SS)}
State evolution equation: \[\begin{bmatrix} x_{k}\\v^{+}_{k}\\ v^{-}_{k}\\ x^{*}_{k} \\ v^{+*}_{k}\\v^{-*}_{k}\end{bmatrix}=\begin{bmatrix} x_{k-1}\\ x_{k-1}v^{+}_{k-1} \\ x^{*}_{k-1}v^{-}_{k-1}\\x^{*}_{k-1}\\ x^{*}_{k-1}v^{+*}_{k-1} \\ x_{k-1}v^{-*}_{k-1}\end{bmatrix} +\mathbf{u}^{a}_{k}\]  \\ \vspace{5pt}
Observation equation: \[\begin{bmatrix}v_{k}\\v^{*}_{k}\end{bmatrix}=\begin{bmatrix}0&1&1&0&0&0\\0&0&0&0&1&1\end{bmatrix}\begin{bmatrix} x_{k}\\v^{+}_{k}\\ v^{-}_{k}\\x^{*}_{k} \\ v^{+*}_{k}\\v^{-*}_{k}\end{bmatrix}+\mathbf{n}^{a}_{k}\] \\  \vspace{5pt}
Estimate of frequency: $\hat{f}_{k}=\frac{1}{2 \pi \Delta T} \Im \left( \text{ln}\left(x_{k} \right) \right)$
\label{Al:N-SS}
\end{algorithm}

\section{Distributed Frequency Estimation}

Among distributed signal processing algorithms, diffusion based algorithms are proven to be suitable for real-time implementation, computationally efficient, and scalable with the size of the network \cite{key-20}-\cite{key-22}. The performance of diffusion based algorithms are dependent  on the average number of connections per-node (average degree) \cite{key-24}; however, power distribution networks are usually sparsely connected \cite{key-29}. Following the approach in \cite{key-21}, we next present a diffusion based distributed Kalman filtering algorithm established on the use of so-called ``bridge nodes'' which is suitable for use in power grids. Consider the standard distributed state space model corresponding  to node $i$ in a network, given in its widely linear form as \cite{key-30}

\begin{equation}
\begin{aligned}
\mathbf{x}^{a}_{k}=&\mathbf{A}^{a}_{k}\mathbf{x}^{a}_{k-1}+\mathbf{u}^{a}_{k}
\\
\mathbf{y}^{a}_{i,k}=&\mathbf{H}^{a}_{i,k}\mathbf{x}^{a}_{k}+\mathbf{n}^{a}_{i,k}
\end{aligned}
\label{eq:distributed state and observation} 
\end{equation}
where $\mathbf{y}^{a}_{i,k}$, $\mathbf{H}^{a}_{i,k}$, and $\mathbf{n}^{a}_{i,k}$ represent the augmented observation vector, augmented observation matrix, and augmented observation noise at node $i$ and time instance $k$.

\begin{figure}
\centering
\includegraphics[scale=0.5, trim= 1cm 0.7cm 1cm 0.5cm]{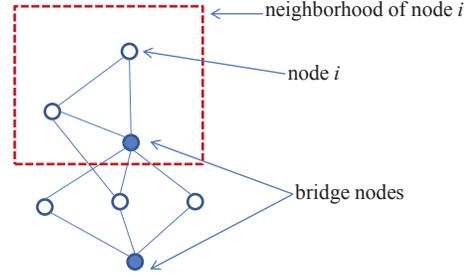}
\caption{Typical network with the selection of bridge nodes denoted by filled-in circles.}
\label{fig:network}
\end{figure}

In the diffusion strategy devised here, nodes of the network, denoted by $\mathcal{N}$, are divided into two sets; bridge nodes, denoted by $\mathcal{B} \subset \mathcal{N}$, and non-bridge nodes. Bridge nodes are selected so that there exists at least one bridge node in the single-hop neighborhood of each non-bridge node and there are no bridge nodes in the single-hop neighborhood of each bridge node. A typical network with its bridge nodes is shown in Figure~\ref{fig:network}; furthermore, a practical algorithm for selecting bridge nodes in a network is presented in \cite{key-31}. 

At the end of every iteration each non-bridge node, $m$, shares its \textit{a posteriori} state estimates, $\hat{\mathbf{x}}^{a}_{m,k|k}$, with the set of its neighboring bridge nodes, $\mathcal{B}_{m}$; then, each bridge node, $i$, diffuses the \textit{a posteriori} estimates of nodes in its neighborhood, denoted by $\mathcal{N}_{i}$, through taking their weighted average, given by
\begin{equation}
\hat{\mathbf{x}}^{a}_{i,k}=\sum_{\forall l \in \mathcal{N}_{i}} \beta_{l,i} \hat{\mathbf{x}}^{a}_{l,k|k}
\label{eq:bridge-diffusion}
\end{equation}
where $\hat{\mathbf{x}}^{a}_{i,k}$ is the diffused state estimate and $\beta_{l,i} \in \mathbb{R}^{+}$ are diffusion coefficients that satisfy $\sum_{\forall l \in \mathcal{N}_{i}}\beta_{l,i}=1$. The bridge nodes share their diffused state estimates with their neighboring nodes; then, each non-bridge node, $m$, diffuses the estimates of its neighboring bridge nodes, $\mathcal{B}_{m}$, in the same fashion that was described for bridge nodes, given by

\begin{equation}
\hat{\mathbf{x}}^{a}_{m,k}=\sum_{\forall l \in \mathcal{B}_{m}} \gamma_{l,m} \hat{\mathbf{x}}^{a}_{i,k}
\label{eq:non-bridge-diffusion}
\end{equation}
where $\gamma_{l,m} \in \mathbb{R}^{+}$ and satisfy $\sum_{\forall l \in \mathcal{B}_{m} }\gamma_{l,m}=1$. The entire process is summarized in Algorithm-\ref{Al:distributed}.

\begin{rem}
Notice that the proposed distributed estimator is only dependent on communication links between bridge nodes and non-bridge, which make it suitable for sparsely connected networks and more robust to link failure.  
\end{rem}

\begin{algorithm}
\caption{Distributed ACEKF} 
Initialize: $\hat{\mathbf{x}}^{a}_{i,0}$, $\hat{\mathbf{y}}^{a}_{i,0}$ and $\hat{\mathbf{M}}^{a}_{i,0|0} \forall i \in \mathcal{N}$ \\ \vspace{5pt}
Estimate $\hat{\mathbf{x}}^{a}_{i,k|k}$ through applying the ACEKF in Algorithm-\ref{Al:ACEKF}.\\ \vspace{5pt}  
If bridge node: calculate $\hat{\mathbf{x}}^{a}_{i,k}$ using (\ref{eq:bridge-diffusion}) and share with nodes in the set $\mathcal{N}_{i}$.\\ \vspace{5pt}
If non-bridge: share $\hat{\mathbf{x}}^{a}_{i,k|k}$ with nodes in the set $\mathcal{B}_{i}$ and calculate $\hat{\mathbf{x}}^{a}_{i,k}$ using (\ref{eq:non-bridge-diffusion}). \label{Al:distributed}
\end{algorithm}

\subsection{Mean error behavior}

Consider the \textit{a posteriori} error given by $\mathbf{e}^{a}_{i,k|k}=\mathbf{x}^{a}_{k}-\hat{\mathbf{x}}^{a}_{i,k|k}$, which can be expressed in terms of the \textit{a priori} state estimate  through
\[
\mathbf{e}^{a}_{i,k|k}=\mathbf{x}^{a}_{k}-\hat{\mathbf{x}}^{a}_{i,k|k-1}-\mathbf{G}^{a}_{i,k}(\mathbf{y}^{a}_{i,k}-\mathbf{H}^{a}_{i,k}\hat{\mathbf{x}}^{a}_{i,k|k-1} ).
\]
Using the observation equation $\mathbf{y}^{a}_{i,k}=\mathbf{H}^{a}_{i,k}\mathbf{x}^{a}_{k}+\mathbf{n}^{a}_{i,k}$ the \textit{a posteriori} error is rearranged to give
\begin{equation}
\mathbf{e}^{a}_{i,k|k}=\left(\mathbf{I}-\mathbf{G}^{a}_{i,k}\mathbf{H}^{a}_{i,k}\right)\mathbf{e}^{a}_{i,k|k-1}-\mathbf{G}^{a}_{i,k}\mathbf{n}^{a}_{i,k}
\label{eq:ape}
\end{equation}
where $\mathbf{I}$ represents the identity matrix (with the same dimensions as $\mathbf{G}^{a}_{i,k}\mathbf{H}^{a}_{i,k}$). Replacing $\mathbf{e}^{a}_{i,k|k-1}=\mathbf{A}^{a}_{i}\mathbf{e}^{a}_{i,k-1}+\mathbf{u}^{a}_{i,k}$ into (\ref{eq:ape}) yields
\begin{equation}
\begin{aligned}
\mathbf{e}^{a}_{i,k|k}=&\left(\mathbf{I}-\mathbf{G}^{a}_{i,k}\mathbf{H}^{a}_{i,k}\right)\mathbf{A}^{a}_{i,k}\mathbf{e}^{a}_{i,k-1}\\
&+\left(\mathbf{I}-\mathbf{G}^{a}_{i,k}\mathbf{H}^{a}_{i,k}\right)\mathbf{u}^{a}_{i,k}-\mathbf{G}^{a}_{i,k}\mathbf{n}^{a}_{i,k}.
\end{aligned}
\label{eq:a posteriori}
\end{equation}

The diffused state estimate error at node $i$ can be expressed in terms of the \textit{a posteriori} error by
\begin{equation}
\begin{aligned}
\mathbf{e}^{a}_{i,k}=&\mathbf{x}^{a}_{k}-\sum_{\forall l \in \mathcal{B}_{i}}\beta_{l,i}\sum_{\forall m \in \mathcal{N}_{i}}\gamma_{m,l}\hat{\mathbf{x}}^{a}_{m,k|k}\\
=&\sum_{\forall l \in \mathcal{B}_{i}}\beta{l,i}\sum_{\forall m \in \mathcal{N}_{i}}\gamma_{m,l}\mathbf{e}^{a}_{m,k|k}.
\end{aligned}
\label{eq:error}
\end{equation}
Replacing (\ref{eq:a posteriori}) into (\ref{eq:error}) yields a recursive expression for the state estimation error as
\begin{equation}
\begin{aligned}
\mathbf{e}^{a}_{i,k} = &\sum_{\forall l \in \mathcal{B}_{i}}\beta{l,i}\sum_{\forall m \in \mathcal{N}_{i}}\gamma_{m,l}\left(\mathbf{I}-\mathbf{G}^{a}_{m,k}\mathbf{H}^{a}_{m,k}\right)\mathbf{A}^{a}_{k}\mathbf{e}^{a}_{m,k-1} \\
&+\sum_{\forall l \in \mathcal{B}_{i}}\beta{l,i}\sum_{\forall m \in \mathcal{N}_{i}}\gamma_{m,l}\left(\mathbf{I}-\mathbf{G}^{a}_{m,k}\mathbf{H}^{a}_{m,k}\right)\mathbf{u}^{a}_{m,k} \\
&\hspace{1em}-\sum_{\forall l \in \mathcal{B}_{i}}\beta{l,i}\sum_{\forall m \in \mathcal{N}_{i}}\gamma_{m,l}\mathbf{G}^{a}_{m,k}\mathbf{n}^{a}_{m,k}.
\end{aligned}
\label{eq:recursive}
\end{equation}
Furthermore, the error can be expressed in terms of the state error covariance matrix estimates by substituting $(\mathbf{I}-\mathbf{G}^{a}_{m,k}\mathbf{H}^{a}_{m,k})=\mathbf{M}^{a}_{m,k|k}(\mathbf{M}^{a}_{m,k|k-1})^{-1}$ into (\ref{eq:recursive})  to give
\begin{equation}
\begin{aligned}
\mathbf{e}^{a}_{i,k}=&\sum_{\forall l \in \mathcal{B}_{i}}\beta_{l,i}\hspace{-0.25em}\sum_{\forall m \in \mathcal{N}_{i}}\gamma_{m,l}\mathbf{M}^{a}_{m,k|k}(\mathbf{M}^{a}_{m,k|k-1})^{-1}\mathbf{A}^{a}_{k}\mathbf{e}^{a}_{m,k-1} \\
&+\sum_{\forall l \in \mathcal{B}_{i}}\beta_{l,i}\sum_{\forall m \in \mathcal{N}_{i}}\gamma_{m,l}\mathbf{M}^{a}_{m,k|k}(\mathbf{M}^{a}_{m,k|k-1})^{-1}\mathbf{u}^{a}_{m,k} \\
&\hspace{1em}-\sum_{\forall l \in \mathcal{B}_{i}}\beta_{l,i}\sum_{\forall m \in \mathcal{N}_{i}}\gamma_{m,l}\mathbf{G}^{a}_{m,k}\mathbf{n}^{a}_{m,k}.
\end{aligned}
\label{eq:recursive-M}
\end{equation} 

Taking the statistical expectation of (\ref{eq:recursive-M}) and assuming that the state evolution and observational noises are zero-mean gives a recursive expression for the mean error as
\[
\begin{aligned}
E&\left[\mathbf{e}^{a}_{i,k}\right]= \\
& \sum_{\forall l \in \mathcal{B}_{i}}\beta_{l,i}\sum_{\forall m \in \mathcal{N}_{i}}\gamma_{m,l}\mathbf{M}^{a}_{m,k|k}(\mathbf{M}^{a}_{m,k|k-1})^{-1}\mathbf{A}^{a}_{k}E\left[\mathbf{e}^{a}_{m,k-1}\right].
\end{aligned}
\]
From the above expression, observe that the mean error at each node is a linear combination of mean errors of different nodes of the network at the previous time instant; therefore, by iteration, the mean error at each node is a linear combination of the mean errors of different nodes of the network at the initial time instant. Thus, if for all nodes in the network $\hat{\mathbf{x}}^{a}_{i,0}$ is an unbiased estimate of $\mathbf{x}^{a}_{0}$ then the algorithm operates in an unbiased fashion.
  
\subsection{Mean square error behavior}

From (\ref{eq:recursive-M}), the meas square error at node $i$ at time instant $k$ can be expressed as
\begin{equation}
\begin{aligned}
\Sigma^{a}_{i,k}=  & E\left[\mathbf{e}^{a}_{i,k}\mathbf{e}^{aH}_{i,k}\right]= \\ &\mathbf{B}_{i,k}\left(\Gamma_{k} \mathcal{E}_{k-1} \Gamma^{H}_{k} + \mathbf{R}\mathcal{U}_{k}\mathbf{R}^{H}_{k}+\mathbf{Q}_{k}\mathcal{G}_{k}\mathbf{Q}^{H}_{k}\right)\mathbf{B}^{H}_{i,k}
\end{aligned}
\label{eq:MSE}
\end{equation}
where
\[
\begin{aligned}
\mathcal{E}_{k} =&E\left[ \left[\mathbf{e}^{aT}_{1,k},...,\mathbf{e}^{aT}_{\left| \mathcal{N}\right|,k} \right ]^{H}\left[\mathbf{e}^{aT}_{1,k},...,\mathbf{e}^{aT}_{\left| \mathcal{N}\right|,k} \right ] \right ]
\vspace{3pt} \\ 
\mathcal{U}_{k} =& E\left[\left[\mathbf{u}^{aT}_{1,k},...,\mathbf{u}^{aT}_{\left|\mathcal{N}\right|,k} \right ]^{H}\left[\mathbf{u}^{aT}_{1,k},...,\mathbf{u}^{aT}_{\left|\mathcal{N}\right|,k} \right ]\right]
\vspace{3pt} \\ 
\mathcal{G}_{k} = &E\left[\left[\mathbf{n}^{aT}_{1,k},...,\mathbf{n}^{aT}_{\left|\mathcal{N}\right|,k} \right ]^{H}\left[\mathbf{n}^{aT}_{1,k},...,\mathbf{n}^{aT}_{\left|\mathcal{N}\right|,k} \right ]\right]
\end{aligned}
\]
which respectively represent the state estimation error, the state noise, and the observation noise augmented cross-covariances between all nodes in the network, whereas
\[
\mathbf{B}_{i,k}=\left[\beta_{i,1}\mathbf{I},...,\beta_{i,\left|\mathcal{B}\right|}\mathbf{I}\right]
\]
with $\mathbf{I}$ denoting an identity matrix (with the same number of rows as the augmented state vector), while the $\left(y,z\right)^{\text{th}}$ element of  $\Gamma_{k}$, $\mathbf{R}_{k}$, and $\mathbf{Q}_{k}$ are given by $\gamma_{y,z}\mathbf{M}^{a}_{y,k|k}(\mathbf{M}^{a}_{y.k|k-1})^{-1}\mathbf{A}^{a}_{y,k}$, $\gamma_{y,z}\mathbf{M}^{a}_{y,k|k}(\mathbf{M}^{a}_{y,k|k-1})^{-1}$, and $\gamma_{y,z}\mathbf{G}^{a}_{z,k}$ respectively. Without loss of generality by assuming that there are no faulty nodes in the network (i.e. all estimates converge to the same state), the state involution and observation matrices are time invariant, and that the state evolution and observation noises are stationary, the covariance matrices $\mathcal{E}_{k}$, $\mathcal{U}_{k}$, and $\mathcal{G}_{k}$ become time invariant and $\Sigma^{a}_{i,k}$ converges. 

From (\ref{eq:error}), $\Sigma^{a}_{i,k}$ can be alternatively expressed as
\[
\Sigma^{a}_{i,k}=\sum_{\forall y \in \mathcal{B}_{i}} \sum_{\forall z \in \mathcal{B}_{i}} \beta_{i,y} \beta_{z,i} \mathcal{V}_{(y,z),k}
\]
where $\mathcal{V}_{(y,z),k}$ is the $(y,z)^{\text{th}}$ element of 
\begin{equation}
\mathcal{V}_k=\Gamma_{k}\mathcal{E}_{k-1}\Gamma^{H}_{k}+ \mathbf{R}_{k}\mathcal{U}_{k}\mathbf{R}^{H}_{k} + \mathbf{Q}_{k}\mathcal{G}_{k}\mathbf{Q}^{H}_{k}
\label{eq:non-bridge-MSE}
\end{equation}
which corresponds to the cross-correlation between the errors of the $y^{\text{th}}$ and $z^{\text{th}}$ nodes. Since $\sum_{\forall y \in \mathcal{B}_{i}}\sum_{\forall z \in \mathcal{B}_{i}} \beta_{i,x}\beta_{y,i}=1$, then $\Sigma^{a}_{i,k}$ is upper-bound by the maximum MSE of its neighboring bridge sensors. 

\begin{rem}
From (\ref{eq:non-bridge-MSE}), observe that the diagonal elements of $\mathcal{V}_{k}$ represent the MSE at each node without the extra layer of diffusion obtained through the bridge nodes (i.e. MSE of conventional diffusion algorithms). Thus, the MSE of the developed algorithm is upper-bounded by the MSE of conventional diffusion algorithms with MSE of both algorithms being equal at bridge nodes, where diffusion takes places only once.
\label{Re:MSE}
\end{rem}

\subsection{Practical implementation}

To the best of my knowledge, all distributed frequency estimators consider every node in the power distribution network to share the same state vector; however, this assumption holds true only if every node in the network is either balanced or experiences the same voltage sag at the same time instant, which is rarely the case as voltage sags can change their characteristics as they propagate throughout the network. This practical consideration can be accommodated for by only sharing the phase increment elements of the state vector. Thus, considering (\ref{eq:widely linear model of v}) and (\ref{eq:distributed state and observation}), the state evolution and observation equations for the distributed frequency estimator become
\begin{equation}
\begin{array}{c}
\underbrace{\begin{bmatrix} x_{k} \\ x^{*}_{k}\end{bmatrix}}_{\mathbf{x}^{a}_{k}}=\underbrace{\begin{bmatrix}1&0\\0&1\end{bmatrix}}_{\mathbf{A}^{a}_{k}}\underbrace{\begin{bmatrix} x_{k-1} \\ x^{*}_{k-1}\end{bmatrix}}_{\mathbf{x}^{a}_{k-1}}+\mathbf{u}^{a}_{k}
\\
\underbrace{\begin{bmatrix} y_{i,k} \\ y^{*}_{i,k}\end{bmatrix}}_{\mathbf{y}^{a}_{i,k}}=\underbrace{\begin{bmatrix}v^{+}_{i,k}&v^{-}_{i,k}\\v^{-*}_{i,k}&v^{+*}_{i,k}\end{bmatrix}}_{\mathbf{H}^{a}_{i,k}}\underbrace{\begin{bmatrix} x_{k} \\ x^{*}_{k}\end{bmatrix}}_{\mathbf{x}^{a}_{k}}+\mathbf{n}^{a}_{k}
\end{array}
\label{eq:new distributed state and observation}
\end{equation}
where the elements of $\mathbf{H}^{a}_{i,k}$ are estimated using Algorithm-\ref{Al:N-SS}. The process is summarized in Algorithm-\ref{Al:DFE}.

\begin{algorithm}
\caption{Distributed frequency estimator (DFE)}
Initialize: $\hat{\mathbf{x}}^{a}_{i,0}$, $\hat{\mathbf{y}}^{a}_{i,0}$ and $\hat{\mathbf{M}}^{a}_{i,0|0} \forall i \in \mathcal{N}$ \\ \vspace{5pt}
For $k=1,2,...$ and $\forall i \in \mathcal{N}$: \\ \vspace{5pt}
Estimate $\mathbf{x}_{i,k}$, $\mathbf{v}^{+}_{i,k}$, and $\mathbf{v}^{-}_{i,k}$ through applying Algorithm-\ref{Al:N-SS}. \\ \vspace{5pt}
Update estimates using the state evolution and observation equations in (\ref{eq:new distributed state and observation}).\\ \vspace{5pt}
If bridge node: calculate $\hat{\mathbf{x}}^{a}_{i,k}$ using (\ref{eq:bridge-diffusion}) and share with nodes in the set $\mathcal{N}_{i}$.\\ \vspace{5pt}
If non-bridge: share $\hat{\mathbf{x}}^{a}_{i,k|k}$ with nodes in the set $\mathcal{B}_{i}$ and calculate $\hat{\mathbf{x}}_{i,k}$ using (\ref{eq:non-bridge-diffusion}). \label{Al:DFE}
\end{algorithm}

\begin{rem}
Note that such an approach is not possible for the WL-SS frequency estimator as all elements of its state vector are tied to the operating conditions of the three-phase power system and the phase incriminating element is not considered in the state vector.
\end{rem}

\section{Simulations}

In this section the performance of the developed frequency estimator is validated and compared to that of the L-SS and WL-SS algorithms in different experiments involving practical power grid scenarios. In all experiments the sampling frequency was $f_{s}=1$~KHz and the voltage measurements were considered to be corrupted by white Gaussian noise with signal to noise ratio of $30$dB.  

\begin{figure}
\centering
\includegraphics[width = 8.8cm,trim = 0cm 1cm 0cm 0cm]{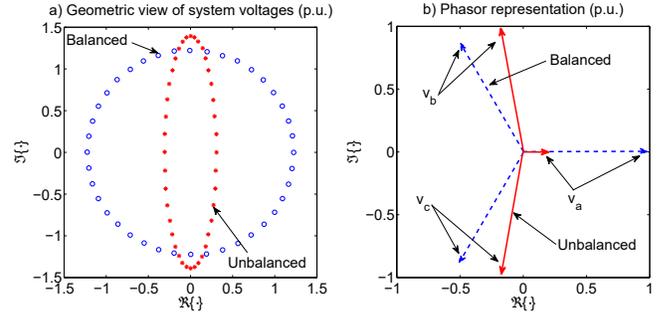}
\caption{System voltages of a three-phase system operating under unbalanced conditions; a) geometric view of the output of the Clarke transform b) phasor representation.}
\label{fig:unbalanced voltages}
\end{figure}

In the first experiment, a balanced system was considered with a fundamental frequency of $f=50$~Hz, which suffers a voltage sag characterized by an $80$\% drop in the amplitude of $v_{a,k}$ and $20$ degree shifts in the phases of $v_{b,k}$ and $v_{c,k}$; furthermore, the frequency of the system experienced a step jump of $2$~Hz. The voltage sag lasted for short duration and the system returned to balanced operating conditions and its nominal frequency once more. The geometric view of the system voltages and the phasor representation of the system are shown in Figure~\ref{fig:unbalanced voltages}.

\begin{figure}
\centering
\includegraphics[width = 8.8cm,trim = 0cm 1cm 0cm 0cm]{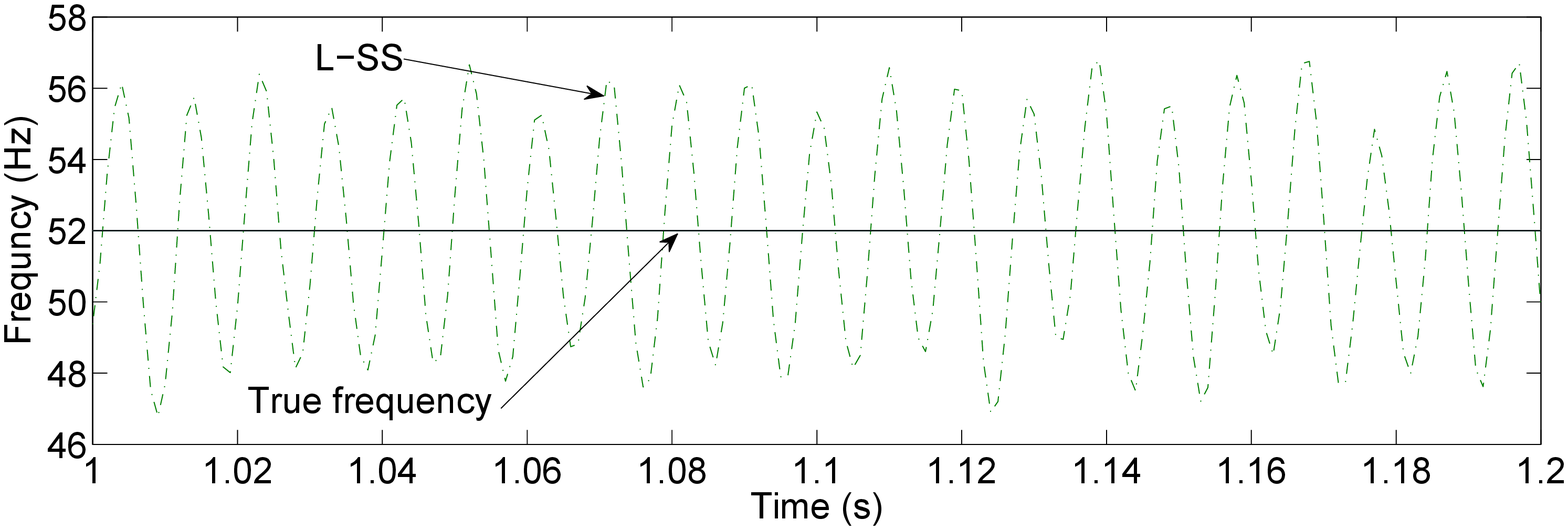}
\caption{Frequency estimation employing the L-SS algorithm for a three-phase power system experiencing a voltage sag.}
\label{fig:L-SS}
\end{figure}

\begin{figure}
\centering
\includegraphics[width = 8.8cm, trim = 0cm -1cm 0cm 0cm]{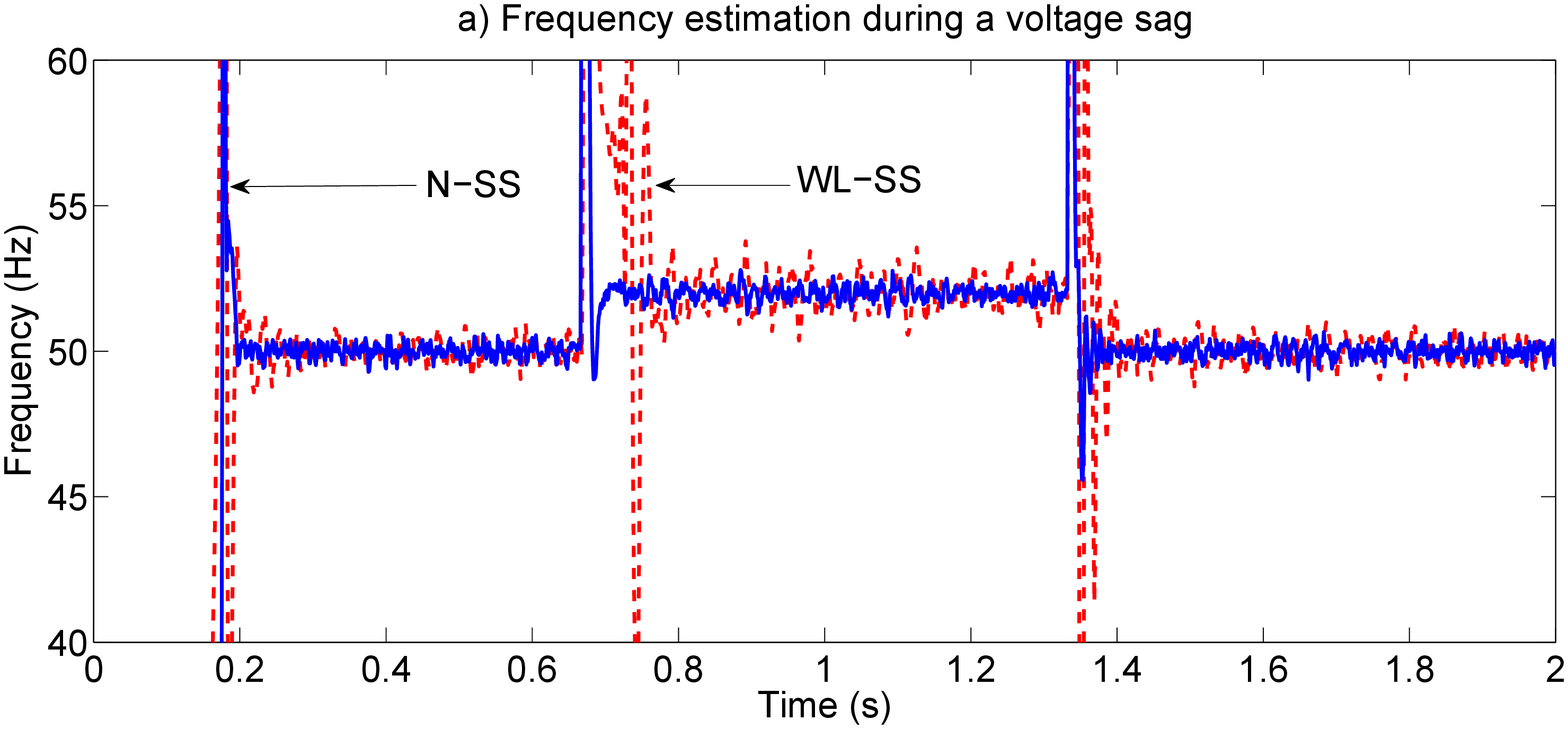}
\includegraphics[width = 8.8cm, trim = 0cm 0.5cm 0cm 0cm]{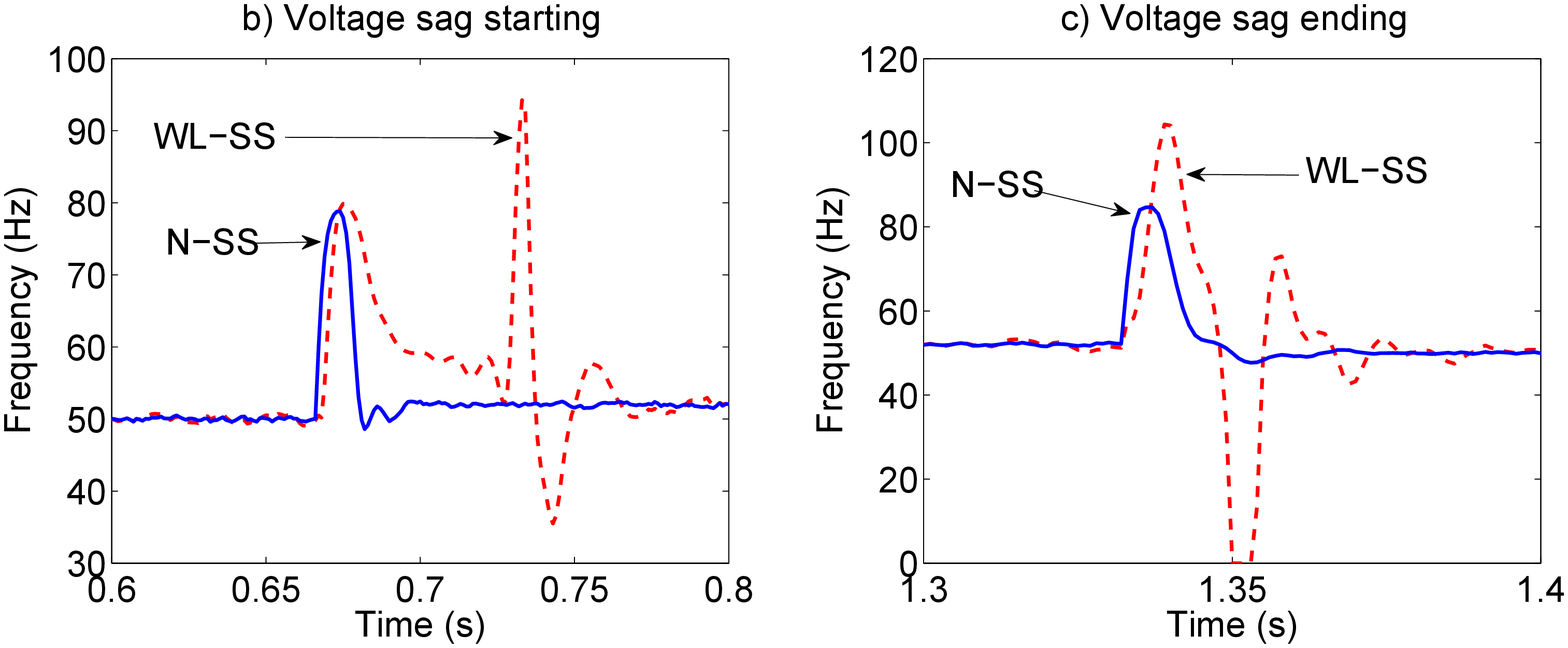}
\caption{Frequency estimation for a three-phase power system experiencing a short voltage sag and a $2$~Hz jump in frequency from $0.667$ to $1.334$ seconds; a) frequency estimation during the voltage sag b) voltage sag starting c) voltage sag ending.}
\label{fig:segment-mixed}
\end{figure}

Figure~\ref{fig:L-SS} shows the estimate of the system frequency obtained by the L-SS algorithm during the voltage sag, which suffered from Large oscillatory errors. The estimates obtained by employing the N-SS and the WL-SS algorithms are shown in Figure~\ref{fig:segment-mixed}. Observe that although N-SS algorithm initially converged at the same time instant as the WL-SS algorithm, when the system was starting to experience the voltage sag, and when the system was recovering from the voltage sag, the N-SS outperformed the WL-SS algorithm in terms of convergence rate; moreover, the N-SS algorithm had a better dynamic behavior (less over and under shoots) and less steady-state variance.

\begin{figure}
\centering
\includegraphics[scale=0.28,trim= 0.5cm 1cm 0cm 1cm]{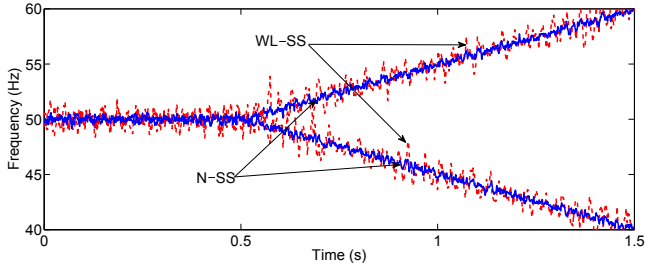}
\caption{Frequency estimation for an unbalanced power experiencing a changing frequency at the rate of $10$~Hz/s.}
\label{fig:unbalancedramp}
\end{figure}

In order to further validate the performance of the N-SS algorithm, in the next experiment, the unbalanced three-phase system characterized in Figure \ref{fig:unbalanced voltages} was considered to experience a rising (\textit{cf}. falling) frequency at a rate of $10$~Hz/s, which typically occurs when power consumption is higher (\textit{cf}. lower) than power generation. The estimates of the system frequency obtained by the N-SS algorithm is compared to that of the WL-SS algorithm in Figure \ref{fig:unbalancedramp}. Notice that the N-SS algorithms accurately tracked the system frequency and outperformed the WL-SS algorithm. 

\begin{figure}
\centering
\includegraphics[width = 8.8cm,trim= 0cm 0cm 0cm 0cm]{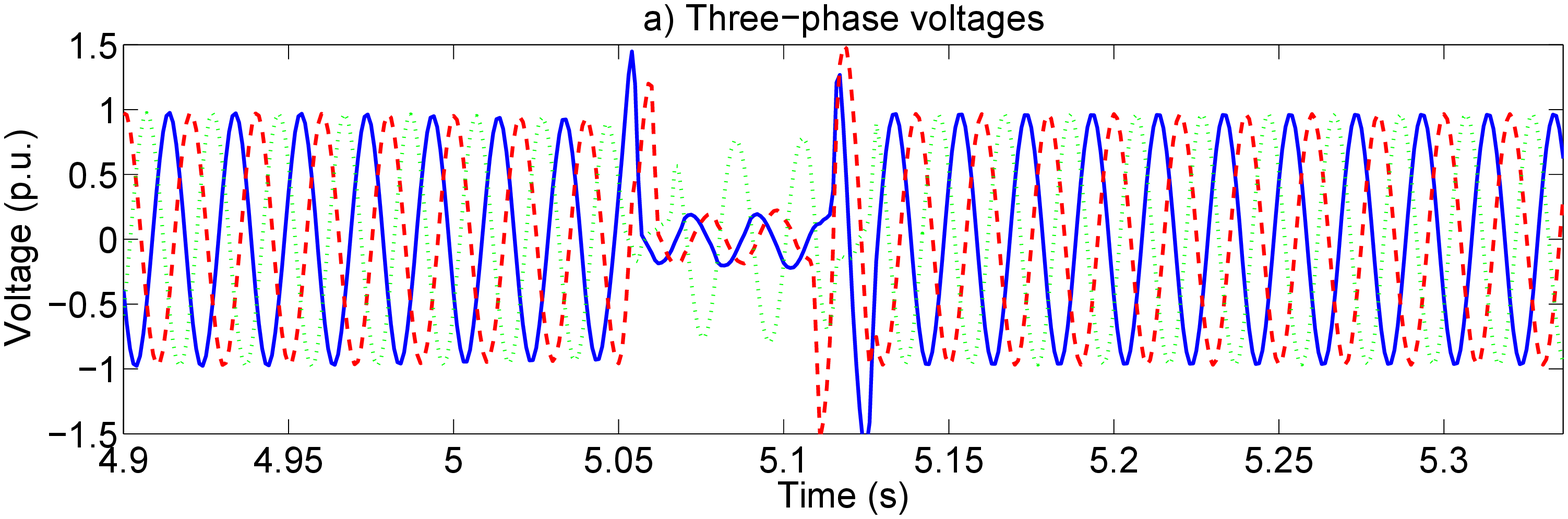}
\\
\includegraphics[width = 8.8cm,trim= 0cm 1cm 0cm 0cm]{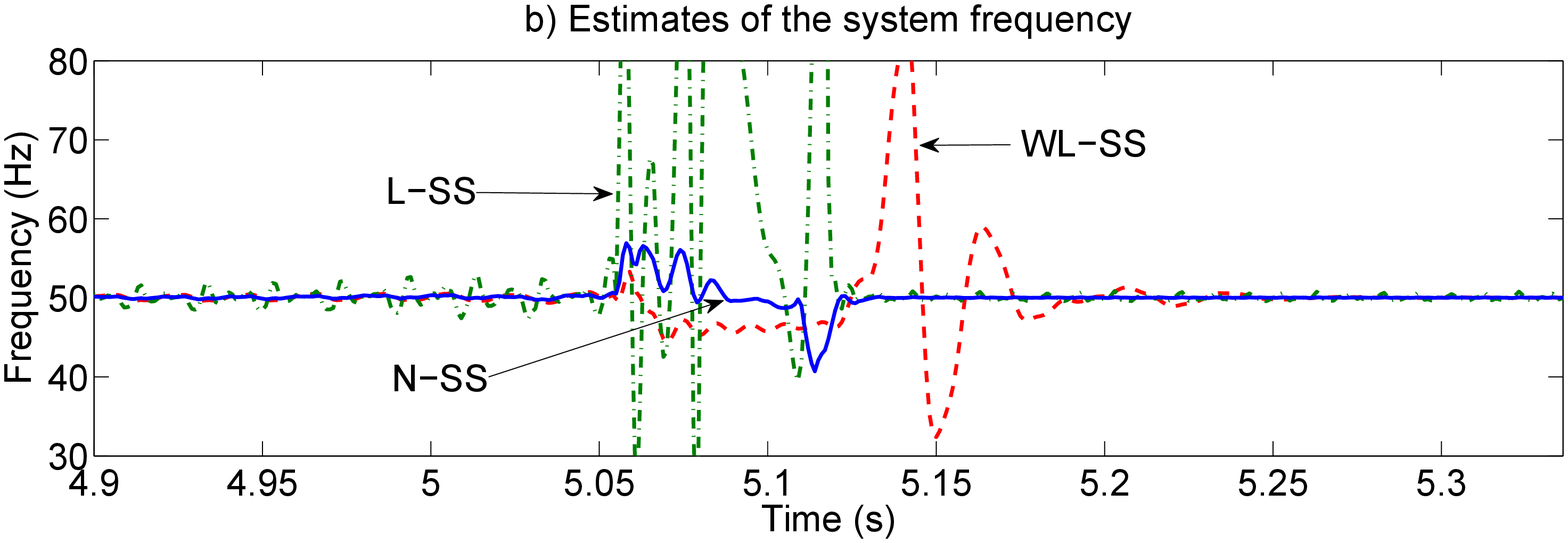}
\caption{Frequency estimation for a real-world three-phase system experiencing a voltage sag: a) three-phase voltages b) estimates of the system frequency.}
\label{fig:recorded1}
\end{figure}

We next examine the performance of the developed algorithms in a more practical setting using data recorded  at a $110/20/10$ k.V. transformer station. The REL $531$ numerical line distance protection terminal, produced by ABB Ltd, was
installed in the station and was used to record the ``phase-ground'' voltages. The three-phase system appearers to be operating in a balanced fashion for the first $5$ seconds, then experiences  a voltage sag at approximately $5.05$ seconds that lasts for around $80$ milliseconds. The recorded data and the estimates of the system frequency are shown in Figure \ref{fig:recorded1}, where from the severely distorted shape of the three-phase voltages during the voltage sag the the presence of higher order harmonics in the system is readily noticeable; furthermore, the following observations can be made:

\begin{enumerate}

\item The L-SS algorithm lost track of the frequency of the system during the voltage sag and was only able converge once the voltage was over. 

\item The WL-SS algorithm tracked the system frequency during the sag; however, the WL-SS algorithm showed unreliable transient behavior when the voltage sag was ending.

\item In comparison to the L-SS and WL-SS algorithms, the N-SS algorithm maintained track of the system frequency and showed outstanding performance. 

\end{enumerate}           

\begin{figure}
\centering
\includegraphics[width = 1\linewidth, trim= 0cm 1cm 0cm 1cm]{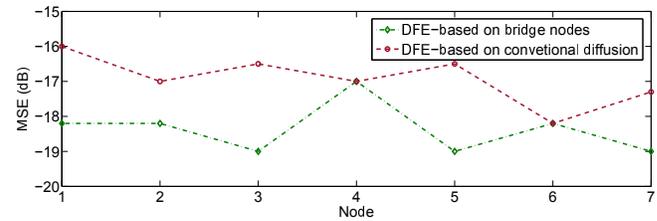}
\caption{Distributed frequency estimation in a network of seven nodes; node $4$ and $6$ correspond to bridge nodes.}
\label{fig:network-balanced}
\end{figure}

We next investigate collaborative frequency estimation in a power distribution network. The network of seven nodes shown in Figure \ref{fig:network} was considered where each node implemented the distributed frequency estimator in Algorithm-\ref{Al:DFE} and all nodes of the network were considered to be operating under balanced conditions. Figure \ref{fig:network-balanced} shows the MSE performance of each node in the network, where it is observed that the MSE of the DFE employing bridge-nodes is upper-bound  by that of the DFE using the conventional diffusion strategy (see Remark~\ref{Re:MSE}). 

\begin{figure}[!h]
\centering
\includegraphics[width = 1 \linewidth, trim = 0cm 1cm 0cm 1cm]{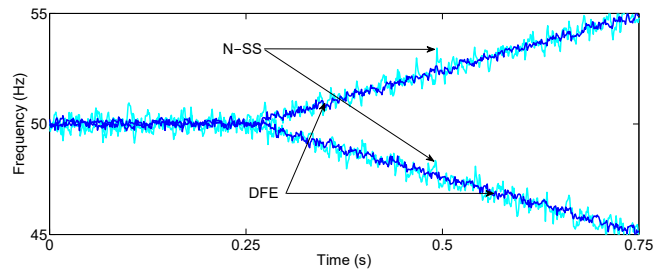}
\caption{Performance of the DFE algorithm: estimates of the non-distributed frequency estimator lay in the light blue region, estimates of the DFE are in dark blue.}
\label{fig:monte}
\end{figure}

We now consider the scenario where only one node of the network operates under balanced conditions. In this experiment, one node of the network was chosen to be operating under balanced conditions while all other nodes where operating under the unbalanced conditions  shown in Figure \ref{fig:unbalanced voltages}; then, Algorithm-\ref{Al:DFE} was implemented to estimate the frequency of the system.  The estimates of the system frequency obtained by employing both the N-SS and DEF algorithms for the node in balanced operating conditions is shown in Figure \ref{fig:monte}. Observe that the DFE  algorithm had a significantly lower steady-state variance and the difference in operating conditions between the nodes of the network did not effect the performance of the DFE algorithm.    

\begin{figure}[!h]
\centering
\includegraphics[width = 1 \linewidth, trim = 0cm 1cm 0cm 0cm]{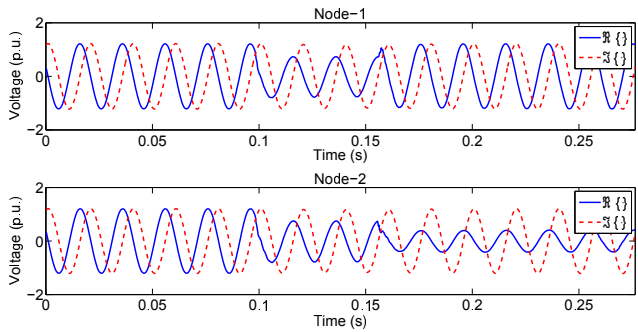}
\caption{Output of the Clarke transform recorded from two adjacent nodes in a power distribution network.}
\label{fig:realdistributeddata}
\end{figure}

\begin{figure}[!h]
\centering
\includegraphics[width = 1 \linewidth, trim = 0cm 1cm 0cm 0.5cm]{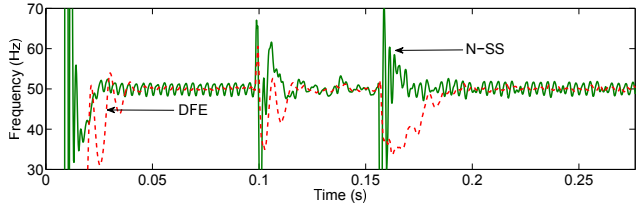}
\caption{Collaborative frequency estimation in a power distribution network; employing the DFE and using real-world data recorded from two adjacent nodes.}
\label{fig:realdistributedfreq}
\end{figure}

Finally, we consider distributed frequency estimation with real-world data recorded from two adjacent nodes in a power distribution network. The recorded data is shown in Figure~\ref{fig:realdistributeddata}, where it is observed that the nodes are experiencing different voltage sag. The estimate of the system frequency employing the N-SS and DFE algorithms are shown in Figure~\ref{fig:realdistributedfreq}. Notice that the DFE had a significantly lower steady-state MSE.

\section{Conclusion}

Frequency estimation in three-phase power systems was revisited and the unified approach for estimating the fundamental frequency of both balanced and unbalanced three-phase systems based on the positive and negative sequenced elements in the output of the Clarke transform presented in \cite{ME} was extended to a distributed setting through introducing its diffusion based distributed dual established on the use of bridge nodes. This was done to address practical issues for implementation of such an frequency estimator in power distribution networks. The analysis indicate that the distributed frequency estimator is unbiased and its MSE is upper bound by that of conventional diffusion algorithms. The algorithm was extensively tested using synthetic and real-world data validating its performance in different operating conditions.   

\balance

\end{document}